\documentclass[aip,apl,twocolumn,groupedaddress]{revtex4}

\usepackage{bm}
\usepackage{graphicx}

\begin{document}

\title{Modulation of Dirac electrons in epitaxial Bi$_2$Se$_3$ ultrathin films on van-der-Waals ferromagnet Cr$_2$Si$_2$Te$_6$}

\author{Takemi Kato,$^{1}$ Katsuaki Sugawara,$^{1,2,3}$ Naohiro Ito,$^{1}$ Kunihiko Yamauchi,$^{4}$ Takumi Sato,$^{1}$ Tamio Oguchi,$^{4}$ Takashi Takahashi,$^{1,2,3}$ Yuki Shiomi,$^{5}$ Eiji Saitoh,$^{2,3,6,7,8}$ and Takafumi Sato$^{1,2,3}$}

\affiliation{$^1$Department of Physics, Tohoku University, Sendai 980-8578, Japan}
\affiliation{$^2$Center for Spintronics Research Network, Tohoku University, Sendai 980-8577, Japan}
\affiliation{$^3$WPI Research Center, Advanced Institute for Materials Research, Tohoku University, Sendai 980-8577, Japan}
\affiliation{$^4$Institute of Scientific and Industrial Research, Osaka University, Ibaraki, Osaka 567-0047, Japan}
\affiliation{$^5$Department of Basic Science, The University of Tokyo, Meguro, Tokyo 152-8902, Japan}
\affiliation{$^6$Institute for Materials Research, Tohoku University, Sendai 980-8577, Japan}
\affiliation{$^7$Department of Applied Physics, The University of Tokyo, Tokyo 113-8656, Japan}
\affiliation{$^8$Advanced Science Research Center, Japan Atomic Energy Agency, Tokai, 319-1195, Japan}

\date{\today}

\begin{abstract}

We investigated the Dirac-cone state and its modulation when an ultrathin film of topological insulator Bi$_2$Se$_3$ was epitaxially grown on a van-der-Waals ferromagnet Cr$_2$Si$_2$Te$_6$ (CST) by angle-resolved photoemission spectroscopy.  We observed a gapless Dirac-cone surface state in 6 quintuple-layer (QL) Bi$_2$Se$_3$ on CST, whereas the Dirac cone exhibits a gap of 0.37 eV in 2QL counterpart. Intriguingly, this gap is much larger than those for Bi$_2$Se$_3$ films on Si(111). We also revealed no discernible change in the gap magnitude across the ferromagnetic transition of CST, suggesting the very small characteristic length and energy scale of the magnetic proximity effect. The present results suggest a crucial role of interfacial coupling for modulating Dirac electrons in topological-insulator hybrids.

\end{abstract}

\maketitle

\section{Introduction}
\vspace{-0.2cm}
One of exciting challenges in the research of topological insulators (TIs) is realization of exotic topological phenomena associated with the time-reversal-symmetry (TRS) breaking, as exemplified by the quantum anomalous Hall effect (QAHE) \cite{Chang1, Checkelsky, Chang2}, the topological magnetoelectric effect \cite{Qi1, Wang, Morimoto}, the magnetic monopole \cite{Qi2}, and the inverse spin-galvanic effect \cite{Garate}. A well-known strategy to realize such fascinating topological phenomena is introduction of magnetic impurities into TI crystals. In fact, this method was successful in realizing the QAHE at low temperatures \cite{Chang1, Checkelsky, Chang2}, whereas disorders are inevitably introduced into the crystal and hinder further enhancement of the observation temperature. 

Magnetic proximity effect is a useful approach to solve this problem. The concept of magnetic proximity effect has been proposed in the stripe phase of high-$T_{\rm c}$ cuprates \cite{Emery} wherein topological defects associated with the formation of stripes are characterized by the spin-gap opening due to the magnetic proximity effect from neighboring antiferromagnetic regions. Recently, ferromagnetic proximity effect (FPE) occurring in a hybrid of ferromagnetic insulator and TI is attracting particular attention as an alternative and likely more promising approach to explore topological phenomena at high temperatures, because it can avoid incorporation of disorders into the TI crystal and reduce strong spin scatterings.  The FPE in a TI hybrid shares a similar framework with the magnetic proximity effect in cuprates in the sense that the former (the latter) is characterized by the magnetic-gap (spin- gap) opening at the metallic topological bands (defects) associated with the proximity effect from neighboring magnetic regions. In TI hybrids, the coupling of ferromagnetic insulator EuS and prototypical TI Bi$_2$Se$_3$ was demonstrated to be useful for stabilizing and even enhancing the ferromagnetism at interface \cite{Katmis}. A heterostructure of TIs, wherein magnetic ions (Cr) are modulation-doped only in the vicinity of top and bottom (Bi,Sb)$_2$Te$_3$ surfaces, was reported to host the axion-insulator phase, triggering further investigations of this exotic phase in intrinsic magnetic TIs such as MnBi$_2$Te$_4$ \cite{Otrocov,Hao,Wu}. The FPE is similar to the well-known superconducting proximity effect in the sense that both can add new (originally absent) properties to the attached metallic layer. On the other hand, the latter is characterized by the penetration of electrons (Cooper pairs) from the superconductor to the attached metallic layer through the interface, whereas the former does not have to involve the penetration of electrons (electron spins) across the interface because the exchange field of ferromagnet can influence the attached metallic layer.
 
To effectively utilize the FPE for realizing novel topological quantum phenomena, a good junction between ferromagnetic insulator and TI is essential. In this regard, a van-der-Waals ferromagnet is a suitable material, since the weak van-der-Waals coupling between adjacent layers would be effective for the atomically flat surface, absence of dangling bonds, and stability against air, all of which are useful for promoting the FPE. However, such hybrid systems involving van-der-Waals ferromagnets and TIs have been scarcely explored \cite{Mogi2, Ji, Yao}.

Amongst several van-der-Waals ferromagnets, we have deliberately chosen Cr$_2$Si$_2$Te$_6$ (CST; $T_{\rm C}$ $\sim$ 31 K) \cite{Carteaux, Ouvrard, Ito} as a template to contact with TI [Bi$_2$Se$_3$ in this work; see Fig. 1(a)], by taking into account the good lattice and symmetry matching as well as its perpendicular ferromagnetism - a prerequisite for realizing the Dirac-cone states with an energy gap called Chern gap or a magnetic gap caused by the time-reversal-symmetry breaking. Recently, an anomalous Hall effect has been reported in a system consisting of 9 quintuple-layers (QL) of (Bi,Sb)$_2$Te$_3$ sandwiched by Cr$_2$Ge$_2$Te$_6$ (a sister compound of CST). While this observation was interpreted in terms of formation of an exchange gap in the Dirac-cone interface states due to the FPE \cite{Mogi2}, it is unclear how strong the FPE influences the Dirac-cone states. In a broader perspective, it is unknown to what extent the interfacial coupling modifies the electronic states of TI/ferromagnetic-insulator hybrid and how such modification is linked to the physical properties. It is important to address these questions from the electronic-state point of view for effectively utilizing the FPE in TIs and its related applications. 

In this article, we report an epitaxial growth of Bi$_2$Se$_3$ thin films on CST bulk single crystal by molecular beam epitaxy (MBE). By visualizing the electronic structure with $in$-$situ$ angle-resolved photoemission spectroscopy (ARPES), we observed a large energy gap in the Dirac-cone state for 2QL Bi$_2$Se$_3$ on CST. The gap magnitude is markedly enhanced compared to that for a Bi$_2$Se$_3$ film on Si(111) and insensitive to the ferromagnetic transition. We discuss implications of this finding in terms of the lattice-strain effect, the interfacial coupling of electronic states, and the FPE.

\section{Experiment and calculation}
\vspace{-0.4cm}
Bulk single crystal of CST was grown by the self-flux method \cite{Ito}. To fabricate a Bi$_2$Se$_3$ thin film, a clean surface of CST was obtained by cleaving with a scotch tape in an ultrahigh vacuum of 3$\times$10$^{-10}$ Torr. A Bi$_2$Se$_3$ film was epitaxially grown by evaporating bismuth (Bi) on the CST substrate kept at $\sim$180 $^\circ$C in selenium (Se) atmosphere of 2.5$\times$10$^{-9}$ Torr in the MBE chamber with the base pressure of 5$\times$10$^{-10}$ Torr. The film was subsequently annealed at $\sim$180 $^\circ$C for 30 min to improve the crystallinity and then transferred to the ARPES-measurement chamber without breaking vacuum. To clarify possible substrate dependence of electronic states, a Bi$_2$Se$_3$ film was also fabricated on Si(111). We used an n-type Si(111) wafer with low electrical resistance (As doping, 1-5 m$\Omega\cdot$cm) to accurately control the annealing temperature during the resistive heating and also to avoid the charging effect during the ARPES measurement. A clean 7$\times$7 surface of Si(111) obtained by flash annealing at 1160 $^\circ$C was at first terminated by a $\sqrt{3}$ $\times$ $\sqrt{3}$ $R$30$^\circ$ Bi buffer layer, and then Bi$_2$Se$_3$ film was fabricated on it with keeping the substrate at 200 $^\circ$C. The growth rate of Bi$_2$Se$_3$ was estimated to be 0.14 $\pm$ 0.01 QL/min by using a quartz-oscillator thickness monitor. This value was always fixed irrespective of the substrate (CST or Si) and the film thickness (2 or 6 QL). Evaporation time was controlled with the accuracy of 10 seconds. The estimated film thickness was totally consistent with the band dispersion measured by ARPES which is known to show an obvious difference among 0QL (substrate), 1QL, and 2QL Bi$_2$Se$_3$ films. The growth process was monitored by the reflection high-energy electron diffraction (RHEED). ARPES measurements were carried out using a MBS-A1 electron analyzer with high-flux helium and xenon discharge lamps at Tohoku University. The He-I$\alpha$ and Xe-I$\alpha$ resonance lines ($h\nu$ = 21.218 and 8.437 eV, respectively) were used to excite photoelectrons. The energy and angular resolutions were set to be 16-30 meV and 0.2$^\circ$, respectively. The sample was kept at $T$ = 40 K in an ultrahigh vacuum ( $\textless$ 1$\times$10$^{-10}$ Torr) during ARPES measurements.  The Fermi level ($E_{\rm F}$) of sample was referenced to that of a gold film electrically in contact with the sample substrate.  First-principles band-structure calculations were carried out by the projector augmented wave method implemented in Vienna Ab initio Simulation Package (VASP) code \cite{VASP} with the generalized gradient approximation (GGA) potential \cite{GGA}. We have also calculated the band structure with the GGA+$U$ and Heyd-Scuseria-Ernzerhof (HSE) methods, as detailed in Appendix. After the crystal structure was optimized, the spin-orbit coupling was included self-consistently.

\begin{figure}
\begin{center}
\includegraphics[width=2.5in]{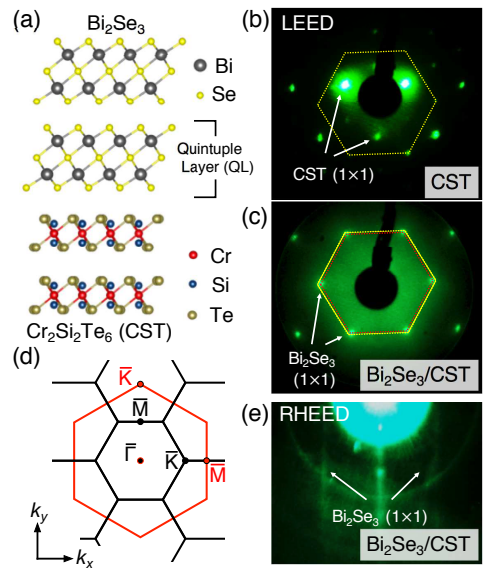}
\caption{(a) Heterostructure consisting of Bi$_2$Se$_3$ (top) and CST (bottom). (b, c) LEED patterns of cleaved CST and 6QL-Bi$_2$Se$_3$/CST, respectively, measured at room temperature with a primary electron energy of 70 eV. Yellow dashed hexagon in (b) traces the $\sqrt{3}$ $\times$ $\sqrt{3}$ $R$30$^\circ$ spots of CST, while a red dashed hexagon in (c) traces the 1$\times$1 spot of Bi$_2$Se$_3$/CST. The yellow hexagon in (b) is superimposed in (c). (d) Projected hexagonal surface BZ of Bi$_2$Se$_3$ (red) and CST (black). (e) RHEED pattern of 6QL-Bi$_2$Se$_3$/CST.}
\end{center}
\end{figure}

\section{Results and discussion}
\vspace{-0.2cm}
First, we present the characterization of Bi$_2$Se$_3$ film on CST. Figure 1(b) shows the low-energy electron diffraction (LEED) pattern of a cleaved surface of CST. The appearance of clear 1$\times$1 spots without additional ones suggests the well-ordered surface as well as the absence of surface reconstruction. When depositing 6QL Bi$_2$Se$_3$ on CST [Fig. 1(c)], the LEED pattern transforms into the 1$\times$1 periodicity of Bi$_2$Se$_3$ lattice. We found that the 1$\times$1 pattern of Bi$_2$Se$_3$/CST is expanded by nearly $\sqrt{3}$ times and rotated by 30$^\circ$ compared to those of pristine CST. A similar behavior was observed for 2QL-Bi$_2$Se$_3$/CST (for details, see Section 1 of Appendix). These results indicate that CST has a surface Brillouin zone (BZ) with a periodicity of nearly $\sqrt{3}$ $\times$ $\sqrt{3}$ $R$30$^\circ$ compared to the 1$\times$1-Bi$_2$Se$_3$/CST, as schematically shown in Fig. 1(d). A careful look at the LEED patterns in Figs. 1(b) and 1(c) further reveals that the $\sqrt{3}$ $\times$ $\sqrt{3}$ $R$30$^\circ$ spots of pristine CST highlighted by a yellow dashed hexagon in Fig. 1(b) do not necessarily coincide perfectly with the 1$\times$1 spots of Bi$_2$Se$_3$/CST [a red dashed hexagon in Fig. 1(c)]. As seen in Fig. 1(c), the red hexagon (Bi$_2$Se$_3$/CST) is slightly smaller than the yellow one (pristine CST). This is consistent with the fact that there is a lattice mismatch of $\sim$ 6 $\%$ between bulk crystals of Bi$_2$Se$_3$ and CST \cite{Carteaux, Ouvrard, Ito} (for details, see Section 5 of Appendix). We have estimated the in-plane lattice constant of Bi$_2$Se$_3$ on CST from the LEED pattern to be 4.1 $\rm{\AA}$, which is in good agreement with that of bulk Bi$_2$Se$_3$ crystal \cite{Vyshnepoisky}. These results suggest a relatively weak lattice strain effect from the CST substrate in 6QL film. As shown in Fig. 1(e), we find sharp streaks in the RHEED pattern, suggesting homogeneous nature of the film. These results support a successful fabrication of high-quality Bi$_2$Se$_3$ film on CST.

Figure 2(a) displays the plot of valence-band ARPES intensity for pristine CST at $T$ = 40 K (in the paramagnetic phase) measured along the $\Gamma$M cut in the CST BZ (i.e. the $\Gamma$K cut in the Bi$_2$Se$_3$ BZ) with the He-I$\alpha$ line ($h\nu$ = 21.218 eV). One can recognize several dispersive bands, e.g., a relatively flat band at the binding energy ($E_{\rm B}$) of $\sim$ 1.8 eV and a dispersive holelike band topped at $E_{\rm B}$ $\sim$ 0.5 eV at the $\Gamma$ point. According to the first-principles band-structure calculations \cite{Kang, ZhangCST}, the former is ascribed to the Cr 3$d$ orbital responsible for the ferromagnetism while the latter to the Te 5$p_{\rm z}$ orbital that forms a valence-band maximum separated by a band gap ($\textgreater$ 0.5 eV) from the conduction-band minimum. As shown in Fig. 2(b), the ARPES intensity undergoes a drastic change upon the formation of 6QL Bi$_2$Se$_3$ film on CST. The flat Cr 3$d$ band at $E_{\rm B}$ $\sim$ 1.8 eV totally disappears and at the same time several highly dispersive Se 4$p$ bands \cite{ZhangBS, Acosta} emerge at $E_{\rm B}$ $\textgreater$ 0.8 eV. Moreover, a new metallic state appears around $E_{\rm F}$ at the $\Gamma$ point. A holelike band topped at 0.5 eV that resembles the Te 5$p_{\rm z}$ band in CST [see Fig. 2(a)] is not ascribed to CST, but to Bi$_2$Se$_3$, because the photoelectron escape depth for the measurement with the He-I$\alpha$ photon ($\sim$ 0.5 nm) is much shorter than the thickness of 6QL Bi$_2$Se$_3$ film ($\sim$ 6 nm). The metallic state near $E_{\rm F}$ is assigned to the conduction band that is pulled down below $E_{\rm F}$ due to the electron doping from Se vacancies, as commonly observed in Se-based TIs \cite{Xia, Sato}. All these spectral features were also observed for the 2QL Bi$_2$Se$_3$ film on CST (see Section C of Appendix for details). The clear dispersive feature of band structure supports the high single crystallinity of our epitaxial film.

To clarify the substrate dependence of the valence-band structure, we fabricated a 6QL Bi$_2$Se$_3$ film on Si(111). A side-by-side comparison of the valence-band structure [Figs. 2(c) and 2(d)] obtained by taking the second derivative of momentum distribution curves (MDCs) reveals an almost identical band dispersion between Bi$_2$Se$_3$/CST and Bi$_2$Se$_3$/Si(111). For example, a bright intensity spot ascribable to the bottom of the ``M''-shaped band around $E_{\rm B}$ $\sim$ 1 eV in Fig. 2(b) is commonly observed at $E_{\rm B}$ = 1.4 eV. A couple of ``$\Lambda$''-shaped energy bands each topped at 0.5 and 1.5 eV, respectively, are also visible in both plots. These results suggest a negligible effect from the substrate to the valence-band dispersion in the case of thick Bi$_2$Se$_3$ films.

\begin{figure}
\begin{center}
\includegraphics[width=2.5in]{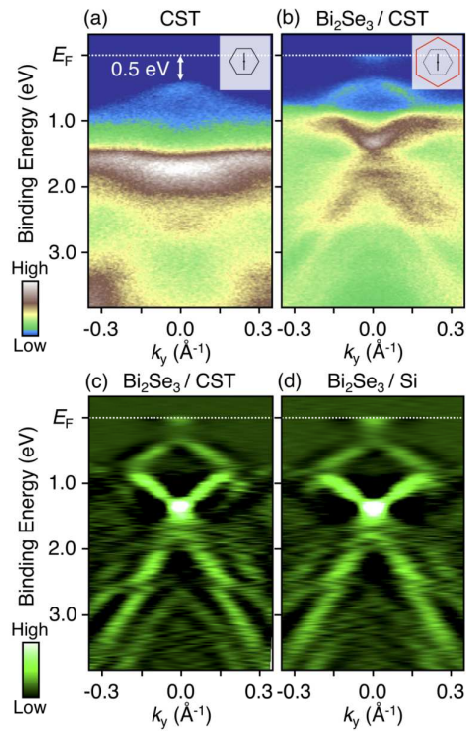}
\caption{(a, b) Valence-band ARPES-intensity of (a) pristine CST and (b) 6QL-Bi$_2$Se$_3$/CST measured at $T$ = 110 K and 40 K, respectively, with the He-I$\alpha$ line ($h\nu$ = 21.218 eV). Inset shows the measured cut (black solid line) in the hexagonal surface BZ. (c, d) Band structures of 6QL-Bi$_2$Se$_3$/CST and 6QL-Bi$_2$Se$_3$/Si(111), respectively, at $T$ = 40 K, obtained by taking the second derivative of MDCs.}
\end{center}
\end{figure}

To elucidate in more detail the electronic states near $E_{\rm F}$, we have performed ARPES measurements with a higher energy resolution using the more bulk sensitive Xe-I photons ($h\nu$ = 8.437 eV). The Xe-I photon is expected to have higher sensitivity to the interface for films thinner than $\sim$ 2QL because the photoelectron escape depth is $\sim$ 2 nm. Figure 3(a) displays the near-$E_{\rm F}$ ARPES-intensity plot around the $\Gamma$ point for 6QL-Bi$_2$Se$_3$/CST measured at $T$ = 45 K (above $T_{\rm C}$ of CST). We find an electronlike feature centered at the $\Gamma$ point within $\sim$ 0.2 eV of $E_{\rm F}$ and an ``M''-shaped band around $\sim$ 0.6 eV. These bands are bulk conduction and valence bands, respectively \cite{Zhangfilm}. We also find an X-shaped Dirac-cone band characterized by touching of V- and $\Lambda$-shaped bands at $E_{\rm B}$ = 0.34 eV at the $\Gamma$ point. A side-by-side comparison of the ARPES intensity between 6QL Bi$_2$Se$_3$ on CST and that on Si(111) in Figs. 3(a) and 3(b) reveals no discernible difference in the near-$E_{\rm F}$ electronic states. A similar behavior is also seen from a comparison of energy distribution curves (EDCs) at the $\Gamma$ point in Fig. 3(c).

In contrast to the thick (6QL) film where the substrate effect to the electronic states is significantly weak, we found that the electronic states suffers an intriguing influence from the substrate in 2QL film. As shown in Fig. 3(d), the ARPES intensity for 2QL-Bi$_2$Se$_3$/CST shows a marked suppression around $E_{\rm B}$ $\sim$ 0.4 eV, obviously different from that for the 6QL film [Fig. 3(a)]. This suggests the opening of an energy gap (Dirac gap) due to the hybridization of wave functions between the top and bottom surfaces \cite{Zhangfilm}. We think that this gap is different from the Chern gap associated with the time-reversal symmetry breaking because the gap persists above $T_{\rm C}$ of CST, as detailed later. One can also recognize a similar gap-opening for 2QL-Bi$_2$Se$_3$/Si(111) as shown in Fig. 3(e), whereas the Dirac cone is obviously shifted downward compared to the case of 2QL-Bi$_2$Se$_3$/CST. We estimated the midpoint of the Dirac gap to be $E_{\rm B}$ = 0.34 and 0.50 eV for 2QL Bi$_2$Se$_3$ on CST and that on Si(111), respectively, by tracing the peak position of EDCs at the $\Gamma$ point [Fig. 3(f)]. This indicates that the 2QL film on Si(111) is more electron doped compared to that on CST. It is expected that the much larger work function of Bi$_2$Se$_3$ (5.6 eV) \cite{Takane} than that of CST (4.49 eV) \cite{Pei} would cause a sizable electron transfer from CST to Bi$_2$Se$_3$, as in the case of Bi$_2$Se$_3$ film on other substrates including Si(111) \cite{Zhangfilm, Hasegawa, Landolt, Chang3}. However, such electron transfer across the interface seems not to take place in 2QL-Bi$_2$Se$_3$/CST, as evident from the experimental fact that the binding energy of the Dirac point for 6QL-Bi$_2$Se$_3$/CST is the same as that of the midpoint of the Dirac gap for 2QL-Bi$_2$Se$_3$/CST (both are at 0.34 eV). It is noted here that a comparison of the work function between CST (4.49 eV) and Si (4.8 eV) suggests higher electron doping in 2QL-Bi$_2$Se$_3$/CST than that in 2QL-Bi$_2$Se$_3$/Si(111), contrary to the experimental results shown in Figs. 3(d) and 3(e). While the exact origin of this behavior is unclear at the moment, we speculate that the presence of Bi $\sqrt{3}\times\sqrt{3}R30^\circ$ buffer structure on the Si surface may reduce the effective work function of Si down to below 4.49 eV.

\begin{figure}
\begin{center}
\includegraphics[width=2.7in]{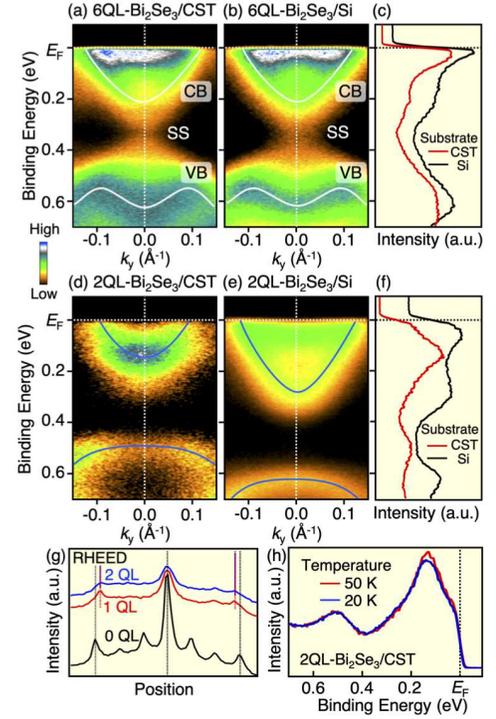}
\caption{(a, b) Near-$E_{\rm F}$ ARPES-intensity plots around the $\overline{\Gamma}$ point for 6QL Bi$_2$Se$_3$ on CST and on Si(111), respectively, measured with the Xe-I$\alpha$ line ($h\nu$ = 8.437 eV) at $T$ = 45 K. White curves are a guide for the eyes to show the energy dispersion of conduction band (CB) and valence band (VB). SS denotes the Dirac-cone surface state. (c) Comparison of the EDCs at the $\overline{\Gamma}$ point for (a) and (b). (d), (e), and (f) Same as (a), (b), and (c) but for 2QL Bi$_2$Se$_3$. Blue curves in (d) and (e) are a guide for the eyes that traces gapped upper and lower Dirac cones. (g) Line profiles of the RHEED patterns for 0QL (pristine CST), 1QL and 2QL-Bi$_2$Se$_3$/CST. Peak positions are indicated by vertical dashed lines. (h) EDC at the $\overline{\Gamma}$ point for 2QL-Bi$_2$Se$_3$/CST measured at $T$ = 50 and 20 K, across $T_{\rm C}$ of CST.}
\end{center}
\end{figure}

By analyzing EDCs in more details [Fig. 3(f)], we found an intriguing difference in the magnitude of the Dirac gap, which is $\sim$ 370 and $\sim$ 300 meV for 2QL-Bi$_2$Se$_3$/CST and 2QL-Bi$_2$Se$_3$/Si(111), respectively. The Dirac gap is enhanced by 70 meV when the substrate is changed from Si to CST. The smaller gap for the Si substrate is not attributed to the experimental artifact such as partial inclusion of 3QL because the gap value of the 3QL film on Si is known to be much smaller ($\sim$150 meV)  \cite{Hasegawa}. There are several possibilities to account for this observation. One of possibilities is a tensile strain from the CST substrate which causes the shrinkage of Bi$_2$Se$_3$ crystal lattice along $c$ axis \cite{Aramberri}, leading to the increase in the hybridization strength of wave functions between the top and bottom surfaces. However, as seen from the RHEED profile for the 0-2QL films on CST in Fig. 3(g), the in-plane lattice constant for the 1QL film immediately relaxes to the bulk value [4.1 $\rm{\AA}$; consistent with the LEED data in Fig. 1(c)] and no further lattice relaxation takes place in thicker films, as evident from the identical peak position between 1- and 2-QL films. Such abrupt lattice relaxation also occurs in the case of Si(111) substrate \cite{Bansal}, causing a lattice mismatch of $\sim$ 8 $\%$ (in both CST and Si cases, the Bi$_2$Se$_3$ lattice is expanded relative to the substrate lattice). The fact that the in-plane lattice constant of both films is already relaxed to the bulk value for $n$ = 1 suggests that there is no effective compressive strain in the Bi$_2$Se$_3$ film and thus it is hard to attribute the gap enhancement to the lattice-strain effect. Another explanation of the gap variation is the interfacial hybridization between the lower Dirac cone in Bi$_2$Se$_3$ and the topmost valence band of CST, because these bands are located in a same ($E_{\rm B}$, ${\bm k}$) region [see Figs. 1(a) and 3(d)]. The presence of interfacial hybridization is supported by our first-principles band-structure calculations for the slab of 2QL-Bi$_2$Se$_3$/CST, as detailed in Section D of Appendix.  These results suggest that the interfacial effect of the band hybridization between Bi$_2$Se$_3$ and CST may be a useful means to modulate the Dirac-cone dispersion.

Although the primary cause for the Dirac gap is thought to be the hybridization between the top/bottom surfaces, one may argue that the gap can be further enhanced by the FPE. However, no obvious change in the gap magnitude across $T_{\rm C}$ [see EDC at the $\Gamma$ point in Fig. 3(h)] seems incompatible with the simple FPE scenario unless the ferromagnetic order is stabilized even above $T_{\rm C}$ of bulk CST (note that no such experiment was reported at the moment). It is inferred that the characteristic length and energy scale of the FPE may be under the detectable limit of the present ARPES measurement. This in return serves as an important guideline for making use of TI/CST hybrids to realize novel topological phenomena, because the accurate tuning of $E_{\rm F}$ within the small magnetic gap and the good crystallinity at the interface comparable to the length scale of the FPE are simultaneously required.

 Finally, we comment on the future research direction based on the present results. While Bi$_2$Se$_3$/CST is not suitable for realizing the QAHE due to the electron doping, it would be useful to fabricate (Bi$_{1-x}$Sb$_x$)$_2$Te$_3$/CST where the Dirac point is situated exactly at $E_{\rm F}$ to examine the possible QAHE at higher temperature. To realize a large magnetic gap by the FPE, the substrate is preferred to have a higher $T_{\rm C}$ and required to have an out-of-plane magnetic easy axis. In this regard, some van-der-Waals ferromagnets such as Cr$_2$Ge$_2$Te$_6$  (a cousin material of CST; $T_{\rm C}$ = 65 K) \cite{Ito}, CrI$_3$  ($T_{\rm C}$ = 61 K) \cite{Huang}, Fe$_3$GeTe$_2$ ($T_{\rm C}$ = 220 K) \cite{Chen}, and Fe-intercalated TiS$_2$ ($T_{\rm C}$ = 111 K) \cite{Takahashi} can be the possible candidates for the ferromagnetic substrate. While we intended to create a Dirac gap below $T_{\rm C}$ of CST by utilizing the spontaneous magnetization along the $c$-axis, it would be also possible to control the magnitude of magnetic gap by applying an external magnetic field and also by controlling the field direction using a vector magnet, as in the case of Kagome magnet \cite{Yin}.

\section{Summary}
\vspace{-0.2cm}
In summary, we investigated the Dirac-cone state and its modulation for ultrathin $n$QL Bi$_2$Se$_3$ ($n$ = 2 and 6) films epitaxially grown on a van-der-Waals ferromagnet Cr$_2$Si$_2$Te$_6$ by angle-resolved photoemission spectroscopy. In contrast to the gapless Dirac-cone state observed in 6QL Bi$_2$Se$_3$ on CST, the 2QL-Bi$_2$Se$_3$ film on CST exhibits a clear energy gap of 0.37 eV at the Dirac point, mainly due to the hybridization between the top and bottom surfaces of Bi$_2$Se$_3$ film. This gap magnitude (0.37 eV) is much larger than that (0.30 eV) for a 2QL-Bi$_2$Se$_3$ film on Si(111), suggesting a strong substrate dependence. We observed no discernible change in the gap magnitude across the ferromagnetic transition of CST, suggesting the very small characteristic length and energy scale of the magnetic proximity effect. The present results lay a foundation for modulating Dirac electrons in TI hybrids with interfacial coupling.

\begin{acknowledgments}

We thank T. Kawakami, T. Taguchi, and Y. Nakata for their assistance in the ARPES experiments. This work was supported by Grant-in-Aid for Scientific Research on Innovative Areas ``Topological Materials Science'' (JSPS KAKENHI Grant numbers JP15H05853, JP18H04215, and JP15K21717), ``J-Physics'' (JP18H04311), JST-CREST (no. JPMJCR18T1), JST-PRESTO (no. JPMJPR18L7), Grant-in-Aid for Scientific Research (JSPS KAKENHI Grant numbers JP17H01139, JP18H01821, JP15H02105, JP26287071, JP19H02424, JP19K22124, JP19H05600, and JP25287079), Research Foundation of the Electrotechnology of Chubu, and World Premier International Research Center, Advanced Institute for Materials Research. N. I. acknowledges support from GP-Spin at Tohoku University.

\end{acknowledgments}

\appendix

\section{Low-energy-electron-diffraction pattern of 2QL-Bi$_2$Se$_3$/CST}
Figure 4(a) shows the low-energy-electron-diffraction (LEED) pattern on the cleaved surface of pristine Cr$_2$Si$_2$Te$_6$ (CST) measured with a primary electron energy $E_{\rm P}$ of 60 eV. One can clearly see only (1 $\times$ 1) spots, similarly to the LEED pattern measured with $E_{\rm P}$ = 70 eV in Fig. 1(b) of main text, supporting the absence of surface reconstruction. As shown in Fig. 4(b), the LEED pattern of 2QL-Bi$_2$Se$_3$/CST is characterized by the 1 $\times$ 1 spot of Bi$_2$Se$_3$ with higher background intensity, similarly to the case of 6QL counterpart in Fig 1(c) of the main text. We speculate that the higher background intensity originates from the surface roughness of Bi$_2$Se$_3$ film due to the finite lattice mismatch between Bi$_2$Se$_3$ and CST.

\begin{figure}
\begin{center}
\includegraphics[width=2.7in]{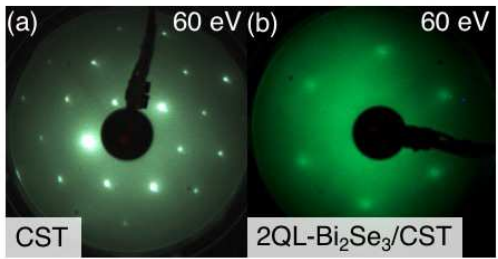}
\caption{LEED patterns on (a) cleaved surface of CST and (b) 2QL-Bi$_2$Se$_3$/CST measured at room temperature with a primary electron energy $E_{\rm P}$ of 60 eV.}
\end{center}
\end{figure}

\section{Atomic arrangement of Bi$_2$Se$_3$ and CST lattices}
We compare in Figs. 5(a) and 5(b) the atomic arrangement of CST and Bi$_2$Se$_3$ , respectively, estimated from the LEED patterns in Figs. 1(b) and 1(c) of the main text. As can be seen from a side-by-side comparison of the unit cell between CST [red diamond in Fig. 5(a)] and Bi$_2$Se$_3$ [yellow diamond in Fig. 5(b)], the unit cell of CST is rotated by 30$^\circ$ and expanded by nearly $\sqrt{3}$ times compared to that of Bi$_2$Se$_3$. When we multiply the in-plane lattice constant of Bi$_2$Se$_3$ (4.1 $\rm{\AA}$; corresponding to the side length of yellow diamond that connects two Se atoms) by exactly $\sqrt{3}$ times, it becomes 7.17 $\rm{\AA}$. This value is slightly larger than the in-plane lattice constant of CST (6.76 $\rm{\AA}$), leading to the $\sim$ 6 $\%$ lattice mismatch between Bi$_2$Se$_3$  and CST at the interface. This difference is better illustrated by directly overlapping the Bi$_2$Se$_3$ and CST lattices in Fig. 5(c).

\begin{figure}
\begin{center}
\includegraphics[width=3.5in]{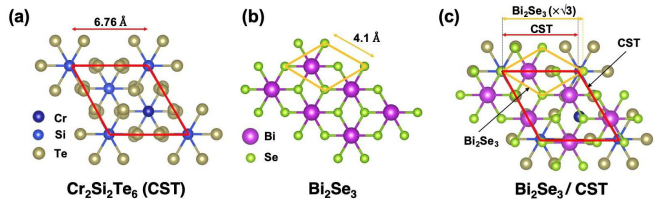}
\caption{(a), (b) Schematics of in-plane atomic arrangement of CST and Bi$_2$Se$_3$, respectively. Red and yellow diamonds show the unit cell of each lattice. (c) Direct overlap of (a) and (b) that highlights the lattice mismatch of $\sim$ 6 $\%$.}
\end{center}
\end{figure}

\section{Comparison of valence-band ARPES intensity between 2QL-Bi$_2$Se$_3$/CST and 2QL-Bi$_2$Se$_3$/Si}

Figure 6 shows a direct comparison of the valence-band ARPES intensity between 2QL-Bi$_2$Se$_3$/CST and 2QL-Bi$_2$Se$_3$/Si(111) measured with the He-I photon. One can recognize the overall similarity in the valence-band dispersion between the two. On the other hand, the spectral feature of 2QL-Bi$_2$Se$_3$/CST appears to be broader and weaker. This is probably because of the smaller CST-substrate size (1 $\times$ 1 mm$^2$) compared to the beam spot of incident light (2-3 mm$\phi$) that causes an increase in the background signal [note that the Si substrate is much larger (2 $\times$ 4 mm$^2$)]. Crystal quality of the two films may be also different in the ultrathin region because the initial growth mode of Bi$_2$Se$_3$ can be different.

\begin{figure}
\begin{center}
\includegraphics[width=2.7in]{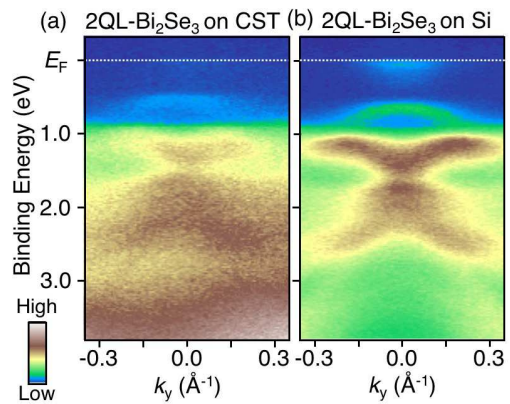}
\caption{Valence-band ARPES intensity for 2QL Bi$_2$Se$_3$ on (a) CST and (b) on Si(111) measured with the He-I$\alpha$ line.}
\end{center}
\end{figure}

\section{Calculated band structure of Bi$_2$Se$_3$/CST}

\begin{figure}
\begin{center}
\includegraphics[width=3.5in]{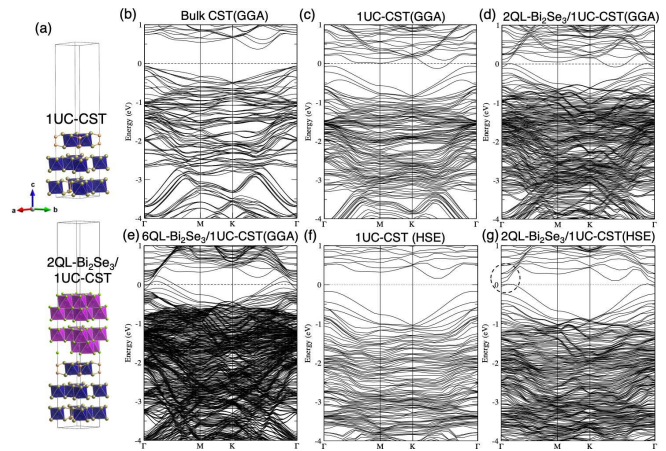}
\caption{(a) Slab structure adopted in the band calculations for (top) 1-unit-cell (UC) CST and (bottom) 2QL-Bi$_2$Se$_3$/1UC-CST. Calculated band structures for (b) bulk CST (GGA), (c) 1UC-CST (GGA), (d) 2QL-Bi$_2$Se$_3$/1UC-CST (GGA), (e) 6QL-Bi$_2$Se$_3$/1UC-CST (GGA), (f) 1UC-CST with the HSE method, and (g) 2QL-Bi$_2$Se$_3$/1UC-CST with the HSE method. Each unit cell (except for bulk CST) includes vacuum layer thicker than 20 $\AA$.}
\end{center}
\end{figure}

To clarify the origin of Dirac gap in 2QL-Bi$_2$Se$_3$/CST, we have carried out the first-principles band-structure calculations for Bi$_2$Se$_3$/CST slabs. While we used bulk (i.e. very thick) CST as a substrate to epitaxially grow Bi$_2$Se$_3$ films, we need to assume a finite thickness of CST in the slab calculation because an infinitely thick lattice cannot be treated in the calculation. We adopted a slab structure containing three unit layers of CST which corresponds to the one unit-cell (1UC) thickness of bulk CST, as shown in the top panel of Fig. 7(a). At first, we calculated the band structure for bulk CST in the ferromagnetic phase. As shown in Fig. 7(b), the calculated topmost valence band shows a holelike dispersion topped at the $\Gamma$ point, separated from the bottom of the conduction band by an indirect band gap of $\sim$0.5 eV, consistent with the ARPES data shown in Fig. 2(a) of the main text. The calculated band structure for 1UC CST in Fig. 7(c) shows a holelike valence band at the $\Gamma$  point as in the case of bulk CST, whereas the conduction bands significantly move downward, resulting in the disappearance of band gap. We have calculated the band structure for 2QL- and 6QL-Bi$_2$Se$_3$ on 1UC-CST by assuming commensurate lattice matching between Bi$_2$Se$_3$and CST [bottom panel of Fig. 7(a) shows the slab structure for 2QL-Bi$_2$Se$_3$/1UC-CST], as shown in Figs. 7(d) and 7(e). One can immediately recognize the presence of many bands in almost entire ($E$, $\textbf{k}$) area including the near-$E_{\rm F}$ region, so that it is hard to draw a meaningful conclusion by directly comparing these calculated bands with the experimental band structure. Such problem is primarily due to (i) the serious underestimation of the bulk-band gap in the slab calculation of 1UC CST and (ii) a large number of atoms involved in a single unit cell of the slab, in particular, for 6QL-Bi$_2$Se$_3$/1UC-CST. Thus, we adopted a different calculation scheme by fixing the slab structure to the smallest unit (2QL-Bi$_2$Se$_3$/1UC-CST) and examined to what extent the slab calculation can qualitatively reproduce the overall experimental feature. To reproduce a finite band gap in 1UC CST, we at first calculated the band structure with the GGA+$U$ method that takes into account the Coulomb energy $U$ of the Cr 3$d$ orbital. However, we were unable to create a band gap even when the $U$ value was increased up to 5 eV. This is consistent with the recent band calculation for a few layer Cr$_2$Ge$_2$Te$_6$ \cite{Menichetti}. Then, by following this literature, we adopted the Heyd-Scuseria-Ernzerhof (HSE06) hybrid functional \cite{Heyd} which mixes the Perdew-Burke-Ernzerhof exchange with 25 $\%$ of the exact nonlocal Hartree-Fock exchange. As a result, we could reasonably reproduce the finite band gap, as shown in Fig. 7(f). Then, we have again calculated the band structure for 2QL-Bi$_2$Se$_3$/1UC-CST by using the HSE method, as shown in Fig. 7(g). One can see from the comparison of Figs. 7(d) and 7(g) that the near-$E_{\rm F}$  band structure becomes much simpler owing to the band-gap opening of CST. A side-by-side comparison of Figs. 7(f) and 7(g) signifies the existence of some bands within the band gap of CST around the $\Gamma$ point in 2QL-Bi$_2$Se$_3$/1UC-CST, as highlighted by a dashed circle in Fig. 7(g). These bands are mainly attributed to the states of Bi$_2$Se$_3$ which correspond to the gapped Dirac-cone bands in the experiment in Fig. 3(d) of the main text. A closer look at the calculation further reveals that there is no band-crossing, supporting the existence of gapless Dirac-cone band. In this context, the calculated band structure is qualitatively consistent with the ARPES data showing the gapped Dirac cone. However, the Dirac gap of 370 meV seen in ARPES for 2QL-Bi$_2$Se$_3$/CST is significantly underestimated in the calculation. This may be because of the assumption of a simplified slab structure with a finite CST thickness and the neglect of a finite ($\sim$6 $\%$) lattice mismatch in the calculation. Although a quantitative comparison of the calculation and experiment is difficult at the moment, the calculation still provides valuable information; the topmost valence band of CST exhibits an energy shift upon attaching 2QL Bi$_2$Se$_3$ on it [Figs. 7(f) and 7(g)]. This supports the band hybridization between Bi$_2$Se$_3$ and CST, in line with the conclusion drawn from the ARPES result.

\bibliographystyle{prsty}

\end{document}